\documentclass[%
a4paper,
prd,
twocolumn,
superscriptaddress,
nofootinbib,
nobibnotes,
amsmath,amssymb,
aps,
fleqn,
floatfix%
]{revtex4-2}
\usepackage{amsfonts,graphics,color}
\usepackage[english]{babel}
\usepackage{graphicx}
\usepackage{float}
\usepackage{dcolumn}
\usepackage{bm}
\usepackage{multirow}
\usepackage{xcolor}
\definecolor{xlinkcolor}{cmyk}{1,1,0,0}
\usepackage[bookmarks=true, pdfnewwindow=true, colorlinks=true, linkcolor=xlinkcolor, citecolor=xlinkcolor, filecolor=xlinkcolor, urlcolor=xlinkcolor, final=true]{hyperref}

\newcommand{\black}{\color{black}}
\newcommand{\blue}{\color{blue}}
\renewcommand{\blue}{\black}

\usepackage[normalem]{ulem}
\definecolor{myblue}{rgb}{0.05,0.1,0.5}

\begin{document}


\title[Carpet--3 event from GRB~221009A]{{\sl Carpet--3} detection of a photon-like air shower with estimated primary energy above 100~TeV in a spatial and temporal coincidence with GRB~221009A }

\author{D.\,D.\,Dzhappuev}
\author{I.\,M.\,Dzaparova}
\affiliation{Institute for Nuclear
Research of the Russian Academy of Sciences, 60th October Anniversary Prospect 7a, Moscow 117312, Russia}
\author{T.\,A.\,Dzhatdoev}
\affiliation{Institute for Nuclear
Research of the Russian Academy of Sciences, 60th October Anniversary Prospect 7a, Moscow 117312, Russia}
\affiliation{Lomonosov Moscow State University, 1-2 Leninskie Gory, Moscow 119991, Russia}
\author{E.\,A.\,Gorbacheva}
\author{I.\,S.\,Karpikov}
\author{M.\,M.\,Khadzhiev}
\author{N.\,F.\,Klimenko}
\author{A.\,U.\,Kudzhaev}
\author{A.\,N.\,Kurenya}
\author{A.\,S.\,Lidvansky}
\author{O.\,I.\,Mikhailova}
\affiliation{Institute for Nuclear
Research of the Russian Academy of Sciences, 60th October Anniversary Prospect 7a, Moscow 117312, Russia}
\author{V.\,B.\,Petkov}
\affiliation{Institute for Nuclear
Research of the Russian Academy of Sciences, 60th October Anniversary Prospect 7a, Moscow 117312, Russia}
\author{E.\,I.\,Podlesnyi}
\affiliation{Norwegian University for Science and Technology (NTNU), Institutt for fysikk, Trondheim, Norway}
\author{N.\,A.\,Pozdnukhov}
\email[Corresponding author; email: ]{nikita.pozdnukhov1@gmail.com}
\affiliation{Institute for Nuclear
Research of the Russian Academy of Sciences, 60th October Anniversary Prospect 7a, Moscow 117312, Russia}
\author{V.\,S.\,Romanenko}
\author{G.\,I.\,Rubtsov}
\affiliation{Institute for Nuclear
Research of the Russian Academy of Sciences, 60th October Anniversary Prospect 7a, Moscow 117312, Russia}
\author{S.\,V.\,Troitsky}
\affiliation{Institute for Nuclear
Research of the Russian Academy of Sciences, 60th October Anniversary Prospect 7a, Moscow 117312, Russia}
\affiliation{Lomonosov Moscow State University, 1-2 Leninskie Gory, Moscow 119991, Russia}
\author{I.\,B.\,Unatlokov}
\affiliation{Institute for Nuclear
Research of the Russian Academy of Sciences, 60th October Anniversary Prospect 7a, Moscow 117312, Russia}
\author{N.\,A.\,Vasiliev}
\affiliation{Lomonosov Moscow State University, 1-2 Leninskie Gory, Moscow 119991, Russia}
\author{A.\,F.\,Yanin}
\author{K.\,V.\,Zhuravleva}
\affiliation{Institute for Nuclear
Research of the Russian Academy of Sciences, 60th October Anniversary Prospect 7a, Moscow 117312, Russia}
\collaboration{Carpet--3 Group}
\noaffiliation


\begin{abstract}
The brightest cosmic gamma-ray burst (GRB) ever detected, GRB~221009A, was accompanied by photons of very high energies. These gamma rays may be used to test both the astrophysical models of the burst and our understanding of long-distance propagation of energetic photons, including potential new-physics effects. Here we present the observation of a photon-like air shower with the estimated primary energy of \blue $300^{+43}_{-38}$~TeV\black, coincident (with the chance probability of \blue $\sim 9\cdot 10^{-3}$\black) with the GRB in its arrival direction and time. Making use of the upgraded Carpet-3 muon detector and new machine learning analysis, we estimate the probability that the primary was hadronic as \blue $\sim 3 \cdot 10^{-4}$\black. This is the highest-energy event ever associated with any GRB.
\end{abstract}

\maketitle


\section{Introduction}
\label{sec:Introduction}
Gamma-ray bursts (GRBs) are intense pulses of soft $\gamma$ and hard X-rays emitted by astrophysical sources (for reviews see e.g.\ Refs.~\cite{Fishman1995,Piran1999,Piran2000,Piran2005,Meszaros2006,Kumar2015,Dai2017}). Typical observed fluences of GRBs are $10^{-7}$~erg/cm$^{2} \lesssim F \lesssim 10^{-4}$~erg/cm$^{2}$, emitted with the characteristic duration of $10^{-2}$~s$\lesssim T_{90} \lesssim 10^{3}$~s encompassing 90\% of the total GRB counts. Two distinct classes of GRBs were identified \cite{Kouveliotou1993,Ghirlanda2009}: short-duration, $T_{90} < 2$~s, GRBs that usually arise in the mergers of compact objects (two neutron stars or a neutron star and a black hole) in binary systems, and long-duration GRBs from the core collapse of massive, $M > 15 M_{\odot}$, stars. 

The prompt GRB emission \cite{Klebesadel1973} is the initial pulse revealing a highly irregular time structure. It is usually followed by an afterglow \cite{Costa1997} that, as a rule, has a smooth light curve extending up to several days. By far the most popular GRB emission scenario invokes the production of the prompt emission in the shock waves internal to the fireball (e.g.\ \cite{Rees1994}; for a recent treatment see e.g. \cite{Rahaman2023}), while the afterglow is believed to be produced in external shock waves (e.g.\  \cite{Rees1992,Meszaros1997,Sari1998}). The afterglow emission covers a wide range of energies from radio to $\gamma$ rays. The lower energy part of the afterglow emission (radio to X rays) is mostly due to synchrotron radiation of electrons accelerated in the external shock wave \cite{Rees1992,Meszaros1993,Paczynski1993,Meszaros1997} while the high-energy part (X rays to TeV $\gamma$ rays) may occur due to inverse Compton scattering of the synchrotron photons on the same electrons \cite{Meszaros1993,Meszaros1994a,Meszaros1994b,Sari2001}. Very high energy (VHE, \mbox{$E > 100$ GeV}) $\gamma$ rays were detected from the afterglows of several GRBs \citep{MAGIC2019,MAGIC2019b,Abdalla2019,HESS2021,Blanch2020}.

On October 9, 2022, the Neil Gehrels Swift Observatory \cite{Williams2023} and the Fermi Gamma-ray Burst Monitor (Fermi-GBM) \cite{Lesage2023} detected an exceptionally bright $\gamma$-ray burst GRB~221009A, sometimes referred to as the brightest of all time (BOAT) \cite{Burns2023}. This long-duration GRB has been detected by numerous instruments in the optical, X-ray and $\gamma$-ray domains \cite{Negro2023,Fulton2023,Kann2023,Laskar2023,Levan2023,Ripa2023,Frederiks2023,Stern2023,Tavani2023,Zheng2024}. The redshift of GRB 221009A was determined as $z = 0.151$ \cite{deUgartePostigo2022,CastroTirado2022,Malesani2023}.

More than 60000 $\gamma$ rays in the energy range 200~GeV -- 7~TeV were recorded with the Large High Altitude Air Shower Observatory (LHAASO) Water Cherenkov Detecor Array (WCDA) during 3000~s after the Fermi-GBM trigger \cite{LHAASO-WCDA-GRB}. The detection of $\gamma$ rays with energies $E > 10$~TeV from GRB 221009A, never observed previously from any other GRB, was reported in \cite{LHAASO-KM2A-GRB}. The LHAASO $1.3\ \rm{km}^2$ Array (KM2A) registered \cite{LHAASO-KM2A-GRB} 142 very high energy photon-like events in the energy range (3--20)~TeV during the time window (230--900)~s after the trigger, while 16.7 were expected from the cosmic-ray background. Nine of them have energies $\gtrsim 10$~TeV in the baseline reconstruction. For each of them, the probability that it is caused by a background cosmic ray lays between 0.045 and 0.17.

The TeV $\gamma$-ray flux from GRB 221009A should be strongly attenuated through the production of electron-positron pairs on extragalactic background light (EBL) photons, $\gamma \gamma \rightarrow e^{+}e^{-}$ \cite{Nikishov1962,Gould1967}. The energy of an individual LHAASO event, $\sim 18$~TeV, first reported in \cite{LHAASO-GCN}, as well as our preliminary announcement \cite{CarpetATel-GRB} of an event with $> 200$ TeV energy, attracted much attention in this context, and numerous new-physics explanations have been discussed. These include mixing of photons with axion-like particles \cite[e.g.][]{Roncadelli-newGRB, ST-GRB-JETPL, Meyer-newGRB, Roncadelli-assessment, EW-axion-and-GRB,axion-DM-and-GRB,axion-2210,Marsh,axion-2304,axion-2305,ST-hostMF-GRB}, Lorentz-invariance violation \cite[e.g.][]{LIV2210-99GeV,LIV-2210-251TeV,LIV-2210-18TeV,LIV-2306-18TeV}, and other exotic scenarios \cite[e.g.][]{other-2210,other-2211-Smirnov,other-2211,other-2212,other-2301,other-2301-Khlopov,other-2309}. However, analyses of published LHAASO results \cite{LHAASO-WCDA-GRB,LHAASO-KM2A-GRB} alone do not require unconventional physics \cite{LHAASO-KM2A-GRB,axion-2310,LIV-2308Piran,LIV-2312Yang}, though may slightly prefer nonzero axion-photon coupling or Lorentz-invariance violation \cite{LHAASO-KM2A-GRB}.

In the present paper, we discuss in detail the Carpet-2 photon-like event associated with GRB~221009A and briefly reported in the telegram \cite{CarpetATel-GRB}. Building upon Ref.~\cite{CarpetATel-GRB}, we leverage data from the newly commissioned large-area muon detector, Carpet-3, which was operational at the time of the event. Dedicated simulations are employed to determine the event's characteristics, including energy and primary particle type, and to assess the probability of a chance association. In Sec.~\ref{sec:data}, we first discuss the experiment and the data set and give references to details of the standard Carpet-2 reconstruction procedure (Sec.~\ref{sec:data:installation}), then present information about the particular event associated with the GRB in Sec.~\ref{sec:data:event}. Section~\ref{sec:anal} presents details of the dedicated analysis of the event, including Monte-Carlo simulations, determination of the primary energy  (Sec.~\ref{sec:anal:MC}) and confronting the event with standard photon selection criteria based on the muon number (Sec.~\ref{sec:anal:mu}). A more advanced classification of the primary particle of the event, based on machine learning, is presented in Sec.~\ref{sec:anal:neural}, see also Appendix~\ref{app:neural}. We proceed in Sec.~\ref{sec:anal:fluence} with the estimate of the effective area of the installation and of the GRB fluence implied by this detection. Section~\ref{sec:results} summarizes our results and puts them in context of the other studies.

\section{Data}
\label{sec:data}

\subsection{Installation and data set} 
\label{sec:data:installation}

Carpet--3 is a ground-based air shower array located at the Baksan Neutrino Observatory of the Institute for Nuclear Research of the Russian Academy of Sciences (Neutrino village, North Caucasus; geographical coordinates 43.273$^\circ$ North, 42.685$^\circ$ East, 1700~m above sea level). The facility includes the central rectangular Carpet array of 400 liquid-scintillator detectors (20$\times$20), each of dimensions of 70$\times$70$\times$30~cm$^3$, providing a continuous area of the array of 196~m$^2$ supplemented by 4 outer stations. In each detector, the energy release is measured with a logarithmic charge-to-digital converter (LQDC), with energy threshold of 8 vertical equivalent muons (VEM). The central Carpet array allows localizing the shower axis and reconstructing the number of relativistic particles $N_{e}$ in the shower, based on the Nishimura-Kamata-Greisen~(NKG) function.

Each of the four outer detector stations (ODS) has 18 detectors (3$\times$6) similar to the ones used in the central Carpet array, all PMT~anode signals from which are summed up. The summed signals from each station are passed to a constant fraction discriminator~(CFD) with a threshold of 0.5 VEM, and then to a time-to-digital converter~(TDC) to measure time delays between the ODS, which are used to determine the arrival direction of the shower (the zenith angle $\theta$ and the azimuthal angle $\phi$).

The underground muon detector (MD) consists of two parallel tunnels, each of which includes 205 plastic scintillator detectors (5$\times$41) with dimensions of 100$\times$100$\times$5~cm$^3$, providing the total MD area of 410~m$^2$. Historically, the muon detector was equipped with detectors in several stages. Initially, in the Carpet--2 experiment, its area was 175~m$^2$ and charge-digital converters~(QDC) measured the total energy release in this area and estimated the total number $n_\mu^{175}$ of detected muons. In 2022, the modernization of MD for the Carpet--3 experiment was completed, as a result of which 235 detectors were added to the MD. Each of these detectors is equipped with a CFD, with a threshold of 0.5~VEM. Thus, the number $n_\mu^{235}$ of muons in the new MD is estimated based on the number of triggered detectors without measuring the energy release in them. The difference in the construction of the two parts of the MD is taken into account in the Monte Carlo simulations of the installation.

\begin{figure}
\includegraphics[width=\columnwidth]{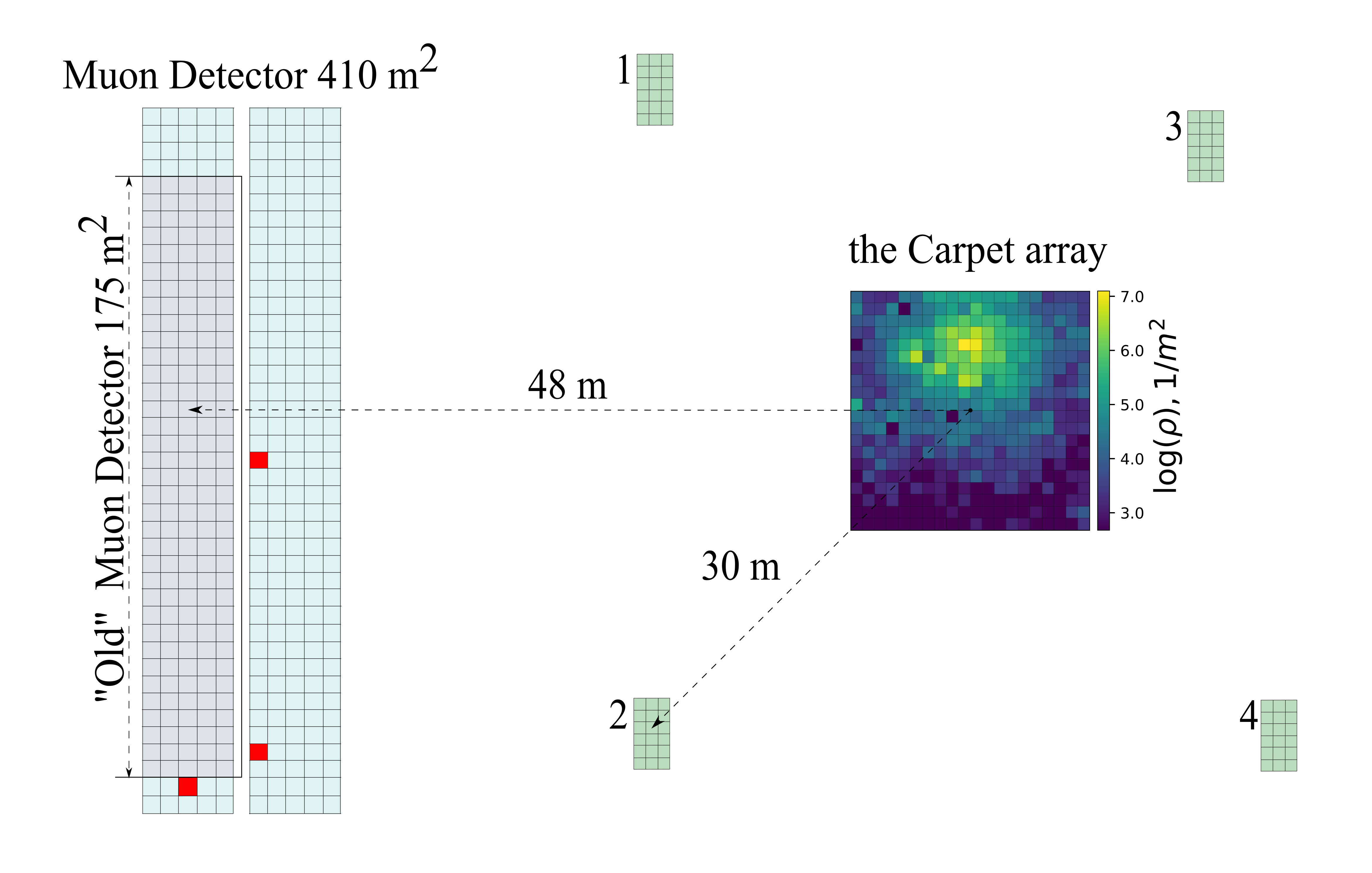}
\caption{
\label{fig:dector-event}
Layout (plot to scale) of the Carpet-3 EAS array and the display of the event associated with GRB~221009A.
The central Carpet consists of 400~liquid-scintillator detectors (20$\times$20) with a total area of 196~m$^2$. The color shows the relative energy release in each detector in a logarithmic scale.
Outer detector stations (ODS) (light green) consist of 18 detectors (3$\times$6) based on a liquid scintillator that are used to calculate the arrival direction of a shower.
The underground muon detector (MD) includes 410 detectors (two tunnels with 5$\times$41 in each) based on a plastic scintillator with a total area of 410~m$^2$, consisting of two parts: the old MD part 175~m$^2$ (grayish blue) and the new part 235~m$^2$ (light blue), used here for the first time for an astrophysical analysis. Detectors that triggered with a threshold greater than 0.5~VEM are highlighted in red.
}
\end{figure}

The layout of the installation collecting data at the day of GRB~221009A, as well as the display of the event we are discussing here, are presented in Fig.~\ref{fig:dector-event}.
For more detailed descriptions of the experiment, data, Monte-Carlo simulations and the standard analysis pipeline, see e.g.\ Refs.~\cite{Carpet2-description2007, Carpet2-Szabelski2009, Carpet2-arXiv2015, Carpet2019ICRC...36..808T, Carpet2019TeVPA-point, BaksanEAS_JETP22}.

The standard extensive air shower (EAS) trigger for recording information about an event is formed by a coincidence circuit under the following condition: 
\begin{itemize}
    \item the total energy release in the central Carpet exceeds 15~VEM, 
    \item four ODS's with a threshold of more than 0.5~VEM have triggered.
\end{itemize}
The typical frequency of this trigger is about 1.3~Hz.

We apply the following criteria when reconstructing events:
\begin{itemize}
    \item  the reconstructed shower axis is inside the central Carpet array excluding the detectors located at the boundary of the central array, 
    \item the reconstructed energy release in the central Carpet array exceeds 5000~VEM,
    \item the number of triggered detectors of the central Carpet array exceeds 200,
    \item the reconstructed zenith angle $\theta \le 40^{\circ}$.
\end{itemize}
All events that satisfy these criteria are included in the data set.
The Carpet--3 data set we use here includes \blue 108468 \black events with data from the 410 m$^{2}$ muon detector recorded between May 20, 2022 and December 31, \blue 2024\black. Due to occasional maintenance, data collection was interrupted a few times. We include only full days of data collection in the analysis, and the number of these days is \blue 667\black. In Appendix~\ref{app:mu}, we discuss, for completeness, the 175~m$^{2}$ MD data which provide for a worse gamma-hadron separation but a much longer exposure. 

\subsection{The event associated with GRB~221009A} 
\label{sec:data:event}
On October 9, 2022, at 14:32:35 UT, that is 1338~s after the Swift trigger and 4536~s after the Fermi-GBM trigger for GRB~221009A, the installation recorded an event with the reconstructed arrival direction $\alpha=289.5^{\circ}$, $\delta=18.4^{\circ}$ in equatorial (J2000) coordinates. This direction is $1.8^\circ$ away from the GRB position in the sky, that is well within the Carpet--3 directional uncertainty ($4.7^{\circ}$ at the $90\%$~CL). The reconstructed number $N_{e}$ of relativistic particles for this shower is $N_{e}^{\rm ev}=36400$. This event had a zero response in the 175~m$^2$ muon detector, $n_\mu^{175,\, \rm ev}=0$, which indicated a rare photon-like event and invited us to publish the telegram \cite{CarpetATel-GRB}. Here, we present a dedicated study of this event using the full 410~m$^2$ detector data. The full MD registered $n_\mu^{\rm ev}=3$ muons. We will see below in Sec.~\ref{sec:anal:mu} that this is much lower than the typical $n_\mu$ of a shower with $N_e \sim N_e^{\rm ev}$ and indicates a likely gamma-ray origin of the event. The readings of individual detector segments are shown in Fig.~\ref{fig:dector-event}. For convenience, the basic information about the event is collected in Table~\ref{tab:event}.
\begin{table}[]
    \centering
    \begin{tabular}{cc}
    \hline
\rule{0pt}{14pt}
date     &  09.10.2022\\
time (UT) & 14:32:35\\
$T-T_0$ & 4536~s\\
zenith angle & 26.5$^\circ$\\
right ascension & 289.5$^\circ$\\
declination & $+18.4^\circ$\\
distance from GRB & 1.8$^\circ$\\
$N_e$ & 36400\\
energy & $\blue 300^{+43}_{-38}$\black~TeV\\
$n_\mu^{\rm 175}$&0 \\
$n_\mu$& 3\\
coincidence&\multirow{2}{*}{$\blue 9.0 \cdot 10^{-3}$}\\
probability& \\
primary hadron&\multirow{2}{*}{$\blue 3 \cdot 10^{-4}$}\\
probability& \\[4pt]
    \hline
    \end{tabular}
    \caption{
    \label{tab:event}
    Properties of the Carpet--3 event associated with GRB~221009A. See the text for details.
    }
\end{table}

Before proceeding with a detailed analysis, we briefly estimate the probability that a background event coincided with the GRB by chance.
The main background for gamma-ray detection with air-shower experiments comes from cosmic-ray events, which can occasionally be muon-poor due to fluctuations. However, the event of interest arrived from the direction close to the Galactic plane, having the Galactic latitude $b\approx 4^\circ$. Galactic sources have been identified as emitters of gamma rays with energies above 100~TeV \cite{Tibet-source, HAWC-sources, LHAASO-sources}; in particular, the unidentified source 3HWC~J1928$+$178, possibly associated with LHAASO~J1929$+$1745, is located in 2.5$^\circ$ from the best-fit arrival direction of the Carpet--3 event \cite{HAWCATel-Carpet-sources}. In addition, the diffuse Galactic-plane gamma-ray emission above 100~TeV has been detected \cite{Tibet-diffuse,LHAASO-diffuse}. Though the sensitivity of present Carpet--3 analyses is insufficient to detect this source, nor to observe the Galactic-plane diffuse flux, occasional Galactic photons may contribute to the background for the GRB observation. The fluxes and spectra of the sources, and of the diffuse background in a given direction, are however known with large uncertainties, which preclude one from using them to model the background. Therefore, we choose to use a conservative data-driven estimate of the background, based on the actual rate of high-energy photon candidate events from the direction in the sky consistent with the GRB up to the Carpet--3 pointing accuracy. 

Due to the Earth's rotation, a given direction in the sky is within the installation field of view for a certain period of time every day, under continuously changing zenith angle. One day is therefore a natural unit of observational time. We determine a high-energy photon-like event as one with reconstructed $N_e \ge N_e^{\rm{ev}}$ and $n_\mu \le n_\mu^{\rm{ev}}$. In \blue 667 \black live days in the data set, \blue 2 \black such photon-like events (including the event associated with GRB~221009A) were detected from the circular area in the sky centered at the GRB direction in equatorial coordinates and having angular radius of Carpet--3 90\%~CL angular resolution.  The Poisson probability of registration of an event satisfying these criteria on the day of GRB~221009A is \blue $3.0 \cdot 10^{-3}$\black, and it does not require trial corrections. Note that the pre-trial p-value of $1.2\times10^{-4}$ reported in the Carpet--2 telegram \cite{CarpetATel-GRB} was obtained in a similar way but (i)~using the time interval of 4536~s instead of one day, (ii)~with slightly stricter criteria for selection of high-energy photon candidates, and (iii)~without the 410 m$^2$ MD information. 

It is important to note that, by construction, the data-driven background, used for the estimate of the probability of a chance coincidence, includes all muon-poor events, no matter if they were caused by unusual cosmic rays or by gamma rays from other sources.

\section{Analysis}
\label{sec:anal}
\subsection{Monte-Carlo simulations and the primary energy}
\label{sec:anal:MC}
Reconstruction of the properties of the primary particle requires Monte-Carlo (MC) simulations. The standard Carpet--2 MC procedure was described in~\cite{Carpet2019JETPLneutrino-old}. The response of the detector to artificial showers is modeled with a dedicated code and stored in the same format as the experimental data are stored. The reconstruction of the MC events is performed by the same procedure and codes for both real and simulated data. 

To analyze the event of interest in more detail, we have generated 6750 air showers caused by primary photons and 6750 showers caused by protons with energies between 100~TeV and 1000~TeV. We use the CORSIKA 7.7420 \cite{Heck:1998vt} EAS simulation package with QGSJET-II-04 \cite{Ostapchenko:2010vb} as the high-energy hadronic interaction model and FLUKA2011.2c \cite{Fluka} as the low-energy hadronic interaction model. All showers are modeled in the energy range 100-1000 TeV with a  spectrum $dN/dE_{0}\propto E_{0}^{-2}$.  The arrival directions of the simulated primary particles are sampled from a two-dimensional Gaussian distribution centered at the initially reconstructed arrival direction of the event of interest, 
with the dispersion reproducing the 90\% CL containment angle of 4.7$^\circ$. The full Monte-Carlo set contains 80609 events.

To estimate the energy of an event associated with GRB~221009A, we use the dependence of $N_{e}$ on the primary energy obtained by fitting the results of the MC simulations for the gamma-ray primaries to a power law. 
This results in the estimate $E_{\gamma}^{\rm ev} =\blue 300^{+43}_{-38}$\black~TeV assuming that the event was caused by a photon.
The full MC simulation data, including the proton primaries, is used in Sec.~\ref{sec:anal:neural}.

\subsection{Muon number}
\label{sec:anal:mu}
We have pointed out that the event associated with the GRB is unusual in terms of its low muon number. We quantify this statement in this section by comparing the observed event with typical air showers of that size detected by Carpet--3 and with simulated gamma-ray induced showers in terms of $n_\mu$.  

The expected distribution of $n_\mu$ for bulk air showers consistent with the GRB-associated event in energy and zenith angle is obtained as follows. We weight the events in real data according to their reconstructed $N_e$ and arrival direction in the horizontal coordinates, azimuth and zenith angles. The weight is calculated as a product of log-normal distribution in $N_e$, centered at $N_e^{\rm{ev}}$ and having the width corresponding to the $\pm 15 \%$ statistical error in $N_e$, and of the two-dimensional Gaussian distribution of directions centered in the observed direction and reproducing the $90 \%$ CL angular resolution of $4.7^\circ$. In the simulation data, the angles of arrival are modeled already taking into account the two-dimensional Gaussian distribution. The model events are weighted taking into account the log-normal distribution of $N_e$, similar to the real data.

\begin{figure}
\includegraphics[width=\columnwidth]{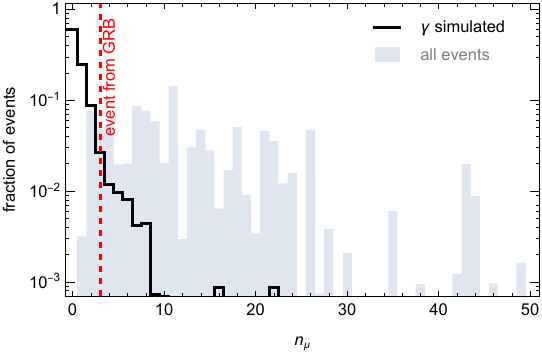}
\caption{
\label{fig:mu_410}
Distribution (PDF) of the number of muons in the Carpet--3 MD. The shaded histogram represents the data; the black histogram represents the simulations assuming gamma-ray primaries; the value $n_\mu=3$ of the GRB-associated event is shown by the vertical dashed line. Event contributions are weighted, as discussed in the text. Note the log scale of the vertical axis.
}
\end{figure}

The obtained distributions are shown in Fig.~\ref{fig:mu_410}).
The fraction of hadronic events that are expected to have $n_\mu \leq 3$ in the $410\ \rm{m}^2$ MD is \blue 0.127 \black.
While the observed event is indeed more photon-like than most of detected EAS, we see that it is not that rare, when the information from the full 410~m$^2$ MD is considered (see Appendix~\ref{app:mu} for the 175~m$^2$ case). In the following Sec.~\ref{sec:anal:neural}, we add more observational information and use a more advanced method to test the photonic origin of the observed event.

\subsection{Neural network gamma-ray classification}
\label{sec:anal:neural}
We estimate the type of primary particle using a neural network classifier trained on the MC event set described above in Sec.~\ref{sec:anal:MC}. The network is trained to distinguish between events with proton and photon primary particles. The signals from 400 scintillator detectors of the central Carpet array are used as a classifier input along with some of the reconstructed parameters, namely, the arrival direction ($\theta, \phi$), the shower axis coordinates at the ground level, $N_e$, the number of muons in the $175\ \rm{m}^2$ detector, the number of muons in the $410\ \rm{m}^2$ detector, and the variable $C_k$ introduced in \cite{Conceicao:2022lkc} that measures the azimuthal nonuniformity of the shower at the ground level. The network itself consists of a convolutional part handling the spatial detector data, which is essentially a 20-by-20 pixel image, and a fully-connected part for the reconstructed parameters. The full architecture of the neural-network classifier is presented in Appendix~\ref{app:neural}.

\begin{figure}
\includegraphics[width=\columnwidth]{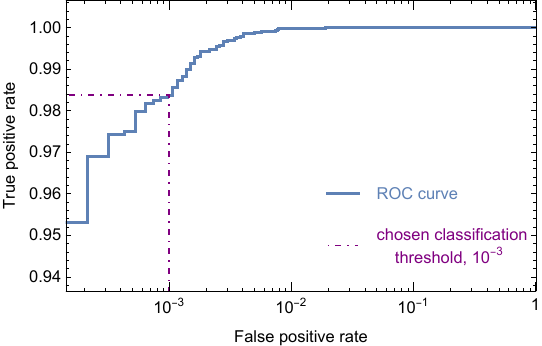}
\caption{
\label{fig:ROC}
ROC curve of the gamma-ray -- proton classifying neural network, evaluated on the test set. The chosen threshold is shown as a dot-dashed line.}
\label{ROC}
\end{figure}

\begin{figure}
\includegraphics[width=\columnwidth]{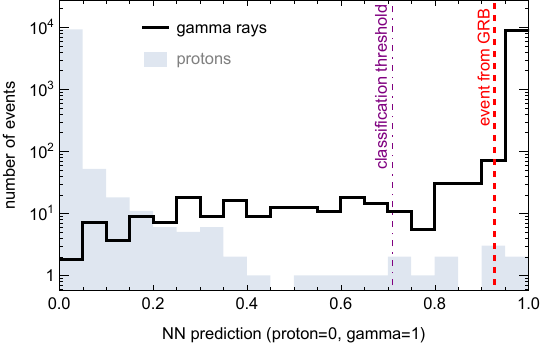}
\caption{
\label{fig:pred_dist}
Predictions distribution of the gamma-ray -- proton classifying neural network. The values of the classification threshold (0.71; dot-dashed line) and of the prediction for the event analyzed in this work (0.927; dashed line) are shown.
}
\label{pred_dist}
\end{figure}

The Monte Carlo set is split into training and test sets, which consist of 62007 and 18602 events, correspondingly. The test set is used for evaluation of the network after training. Fig.~\ref{fig:ROC} 
shows the resulting receiver operating characteristic (ROC) curve. It demonstrates that the network performs well on the MC data, reaching $10^{-3}$ proton background rejection factor without a significant decrease in the photon selection efficiency. Fig.~\ref{fig:pred_dist} 
presents the network prediction distribution. A prediction of 1 corresponds to a photon event, 0 to a proton event. For a chosen background rejection factor of $10^{-3}$, the classification threshold is equal to 0.71. This means that events with prediction values above 0.71 are classified as photons.

We then use the classifier for the event discussed in this work. Data are processed and normalized in the same way as in MC simulations. The resulting prediction value is 0.927, well above the classifier threshold. This gives additional evidence for the photon nature of this event. Based on simulations, the probability that this event is a misclassified hadron is $\approx 3\cdot10^{-4}$. This is calculated as the fraction of MC hadrons that the network incorrectly classifies as photons with the same or higher prediction values.

Within the angular resolution of Carpet--3 of 4.7 degrees, there were \blue 6 \black events in \blue 667 \black live days of data having $N_e \ge N_e^{\rm ev}$ and the same or higher value of neural network prediction, that is, more photon-like, than the GRB-associated event. The probability that a background high-energy photon-like event could arrive on the day of the gamma-ray burst is therefore \blue $9.0 \cdot 10^{-3}$\black.

\subsection{Effective area and fluence estimates}
\label{sec:anal:fluence}
For a fast transient like a GRB, it is not straightforward to compare fluxes measured by different instruments in non-coinciding period of time. A useful quantity to compare is fluence, that is the amount of energy per unit are integrated over the entire time of the burst. To estimate the fluence, one needs some assumptions about the source, in particular, about its spectrum, which we assume here to be a $E^{-2}$ power law. This spectrum is then convolved with the energy-dependent effective area of the installation to obtain the total number of events expected to be detected. Equating this number to the observed one, we obtain the normalization of the assumed spectrum and estimate the fluence. 

\begin{figure}
\includegraphics[width=\columnwidth]{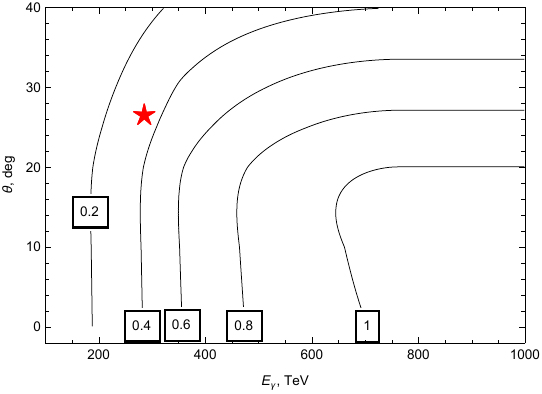}
\caption{
\label{fig:eff}
Contour plot of the photon detection efficiency as a function of energy and zenith angle for the selection cuts adopted in the present work. Numbers in squares represent the efficiency values. The star indicates the values for the GRB-associated event.
}
\end{figure}

The effective area of Carpet--3 is determined by a product of the geometrical area of the installation and the efficiency for photon detection, estimated from Monte-Carlo simulations. For the selection cuts used here, which imply reconstructed shower axes within the central Carpet but without the perimeter detector stations, the geometrical area is $1.6 \times 10^6$~cm$^2$. The efficiency depends on both primary energy $E_\gamma$ and zenith angle $\theta$. To estimate it, we have thrown Monte-Carlo air showers with gamma-ray primaries to the area considerably larger than the installation and find a ratio between the number of events passed all selection criteria and the number of thrown showers with axes within the geometrical area we use. The dependence of efficiency on $E_\gamma$ and $\theta$ is shown in Fig.~\ref{fig:eff}. 
For parameters of the event associated with GRB~221009A, the efficiency is 0.38. Note that this simulation assumed the selection criteria used in the present work, so the values of efficiency plotted in Fig.~\ref{fig:eff} may not be applicable to other analyses. 

The estimated GRB~221009A fluence above 100~TeV is $\mathcal{F}\approx (1.1 \pm 0.9)\times10^{-3}$~erg/cm$^{2}$ (68\%~CL), which is in a reasonable agreement with the values found in the lower energy bands, corrected for absorption. The estimate is based on a single event and thus suffers from large uncertainties, estimated here based on the Poisson statistics as recommended by Ref.~\cite{PDG}. 

\section{Discussion and conclusions}
\label{sec:results}
\subsection{Observational conditions}
\label{sec:results:za}
\begin{figure}
\includegraphics[width=\columnwidth]{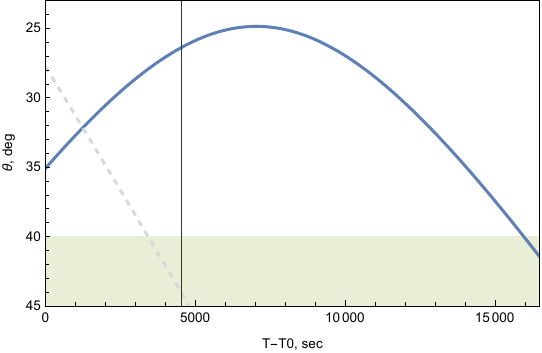}
\caption{
\label{fig:zenith}
Temporal dependence of the zenith angle of the GRB~221009A as seen by Carpet--3 (full blue line). The dashed gray line is the same dependence for LHAASO. The vertical maroon line indicates the Carpet--3 event arrival time. The shaded area at the bottom corresponds to zenith angles $\theta>40^\circ$ excluded in the analysis.
}
\end{figure}

We report on the Carpet--3 observation of a rare photon-like air shower, whose arrival direction and time match those of the exceptional GRB~221009A. Other experiments, including larger ones, have not reported events with those high energies from this, nor from any other, GRB. This may be related to the temporal structure of the very high energy afterglow, combined with visibility conditions for different experiments. Indeed, Carpet--3 observed the event 4536~s after the burst trigger, high above the horizon (see Fig.~\ref{fig:zenith}).
At this moment, the burst site was seen at considerably larger zenith angle at LHAASO, close to the limit of the field of view of the experiment. LHAASO did not report on the GRB~221009A observations beyond 2000~s post trigger. For High-Altitude Water Cerenkov Detector (HAWC), the direction of interest was below the horizon \cite{HAWC-GCN-not-seen}. 

LHAASO observed a hardening of the gamma-ray spectrum with time for GRB~221009A~\cite{LHAASO-WCDA-GRB,LHAASO-KM2A-GRB}. Very high energy, $E>100$~GeV, photons have been detected even days after the trigger both from GRB~221009A by Fermi Large Area Telescope (LAT) \cite{400-GeV-photon,SternTkachev,FermiCollab-GRB} and from GRB~190829A by the High Energy Stereoscopic System (H.E.S.S.) \cite{HESS:190829A}. A quantitative assessment of the development of the afterglow at high energies would be strongly model-dependent.

\subsection{Origin of energetic photons in GRB}
\label{sec:results:GRB}
Let us discuss briefly several possible mechanisms of multi-TeV gamma-ray production in the afterglow of GRB~221009A. Below the observable gamma-ray energy of several TeV, the most natural mechanism of gamma-ray production in GRB afterglows is synchrotron self-Compton (SSC) \cite{Derishev2019,MAGIC2019b}. Above the observed energy of 10~TeV, however, the Klein-Nishina effect typically sets in, resulting in a significant suppression of the interaction rate between the electrons and photons, leading to a downturn in the produced gamma-ray spectrum. 
Therefore, another mechanism is needed to produce the observable $\gamma$-rays with the energy $E > 100$~TeV.

It is hardly possible to explain the detection of the reported event with the proton synchrotron radiation. Indeed, the maximum characteristic energy of the synchrotron photons produced by the accelerating protons is $E_{\rm sp,max} \sim 100$~GeV (in the plasma comoving rest frame) \cite{Kumar2012}. The maximum observable energy of the same photons is $E_{\rm sp,max} \times D / (1+z) \sim 1$~TeV since the value of the Doppler factor $D$ could hardly significantly exceed ten at relatively late time, thousands of seconds after the trigger. We note that the synchrotron-photon energy upper limit of \cite{Kumar2012} does not apply for secondary charged particles, e.g. for electrons and positrons produced in photohadronic or hadronuclear interactions.

Likewise, the photohadronic scenario for the event is strongly disfavoured due to the very low efficiency of the photopion process. As it is commonly done for blazars (for instance, see Ref.~\cite{Dzhatdoev2022}), a very strong upper bound on the efficiency could be set from the apparent absence of strong absorption of GeV-TeV gamma rays in the observable spectrum of GRB~221009A.

Therefore, a nonconventional astrophysical mechanism is required to produce delayed, $\gtrsim$500~s,  gamma rays with the energy of $\sim$10~TeV or higher. One such example was presented in \cite{Dzhatdoev2024}, inspired by previous studies that accounted for interactions of high-energy neutrons escaping from their sources \cite{Berezinskii1977,Eichler1978,Kirk1989,Tkaczyk1994,Atoyan2003,Dermer2004b}. During the GRB prompt emission phase, protons and/or nuclei could be accelerated to high energies, {$E > 1$~PeV/nucleon} in the fireball rest frame. In contrast to the later afterglow phase ($>2000$~s after the Fermi-GBM trigger), during the prompt emission phase the accelerated protons and/or nuclei interact with the dense photon fields inside the fireball relatively efficiently, producing gamma rays, electrons, positrons, neutrinos, and neutrons\footnote{Specific parameters of this scenario may be constrained from non-observation of high-energy neutrinos from GRB~221009A by IceCube \cite{IceCube-GRB221009A} and KM3NeT \cite{KM3NeT-GRB221009A}.}. These neutrons escape from the magnetic fields of the fireball freely and then interact with the interstellar matter of the star-forming region (SFR) where the GRB is located, creating a flux of hyper-relativistic electrons and positrons. These, in turn, radiate synchrotron photons in the magnetic field of the SFR, eventually producing an observable flux of gamma rays at energies $E\sim$10~TeV for typical values of parameters.
For an order-of-magnitude stronger than typical SFR magnetic field, or $\approx$3-4 times higher energy of the accelerated protons, the observable gamma-ray energy of $\sim$100~TeV is achievable in this scenario. The time delay comes from the angular spread of the neutron beam \cite{Dermer2004b}. Thus, the proposed mechanism assumes the presence of a delayed ``echo'' from the prompt emission induced by the escaping and interacting energetic neutrons.

\subsection{Comparison with the LHAASO fluence}
\label{sec:results:fluence}
\begin{figure}
\begin{center}
\includegraphics[width=\columnwidth]{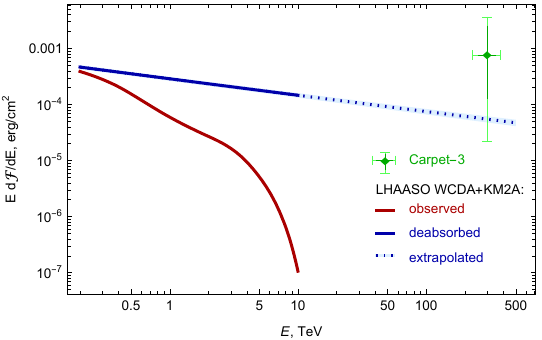}
\end{center}
\caption{
\label{fig:fluence}
Comparison of the photon fluence of GRB~221009A estimated from the Carpet--3 observation (dark green: 68\%~CL, light green: 95\%~CL errors) with the extrapolation of LHAASO results. 
}
\end{figure}

In Ref.~\cite{LHAASO-KM2A-GRB}, LHAASO presented the observed energy spectra, based on the combination of WCDA and KM2A data, and the best power-law fits of the spectrum corrected for absorption within the extragalactic background light model of Ref.~\cite{Saldana-Lopez}. The spectra have been presented for two intervals, 230~s to 300~s and 300~s to 900~s after the trigger. It has also been demonstrated there that the light curve presented in Ref.~\cite{LHAASO-WCDA-GRB} for WCDA fits well the KM2A observations up to the highest energies, thus the energy emitted between 900~s and 2000~s can be estimated. We use this information to obtain the differential fluence, $d\mathcal{F}/dE$, as the function of energy $E$, from the LHAASO measurements. The same quantity, estimated at the energy of the Carpet--3 observed events from the result of Sec.~\ref{sec:anal:fluence}, is in order-of-magnitude agreement with the extrapolation of the \textit{intrinsic} spectrum determined by LHAASO, see Fig.~\ref{fig:fluence}.
\begin{figure}
\includegraphics[width=\columnwidth]{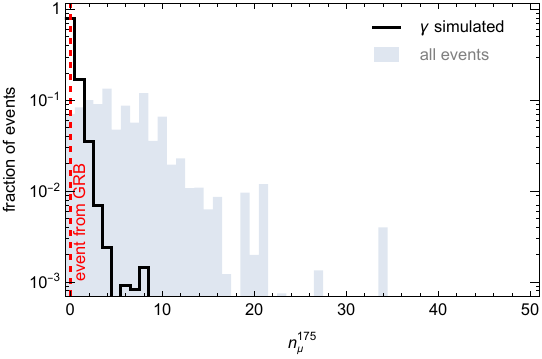}
\caption{
\label{fig:mu_175}
Same as Fig.~\ref{fig:mu_410} but for the 175~m$^{2}$ Carpet--2 MD. 
}
\end{figure}

\subsection{Conclusions}
\label{sec:results:concl}
As we discussed above, gamma rays with the energies of $\sim 300$~TeV cannot reach us from extragalactic sources, provided the Standard Model of particle physics describes correctly the photon propagation. Therefore, if the event reported here is a photon (we estimate the probability of the gamma-ray primary as $1-3\cdot 10^{-4}= 0.9997$) arrived from GRB~221009A (probability $1-\blue9\cdot 10^{-3}=0.991$\black), and reached us from the GRB site, then its detection by Carpet--3 could be seen as a manifestation of new fundamental physics. Relevant scenarios have been explored in the literature discussed in Sec.~\ref{sec:Introduction} and will be scrutinized, with the help of the information about the Carpet--3 event presented here, in future publications. 
The presented results are also relevant for constraining the strength of extragalactic magnetic fields, and for understanding the mechanisms of GRB emission at very high energies.

\section*{Acknowledgements}
We are indebted to L.~Bezrukov, N.~Kalmykov, P.~Satunin and Yu. Stenkin for interesting discussions. This work is supported in the framework of the State project ``Science'' by the Ministry of Science and Higher Education of the Russian Federation under the contract 075-15-2024-541. 

\section*{Data availability}
The data used in this study are available from the corresponding author upon a reasonable request.

\vskip 2mm

\appendix

\section{Carpet--2: the 175 m$^2$ muon detector}
\label{app:mu}
The Carpet--2 data set of April 7, 2018 - December 31, \blue 2024 \black includes \blue 261468 \black events detected with the 175~m$^2$ MD data in \blue 1676 \black live days. The number of events with $N_e>N_e^{\rm ev}$ and $n_\mu=0=n_\mu^{175}$, within 4.7$^{\circ}$ from the GRB~221009A direction, is \blue 7\black, including the event under study. The probability of registering an event that satisfies these criteria on the day of GRB~221009A is \blue $4.2 \cdot 10^{-3}$\black. For completeness, we also provide a conservative estimate that the discussed event may be caused by a hadronic primary using only the old 175~m$^2$ MD data, see Fig.~\ref{fig:mu_175}, 
as \blue 0.070\black. We do not perform the machine-learning analysis for Carpet--2 data here because of lower photon-hadron separation power of the old MD.

\section{The architecture of the neural network}
\label{app:neural}
\begin{figure}[H]
\vskip 1mm
\centering
\includegraphics[width=0.95\columnwidth]{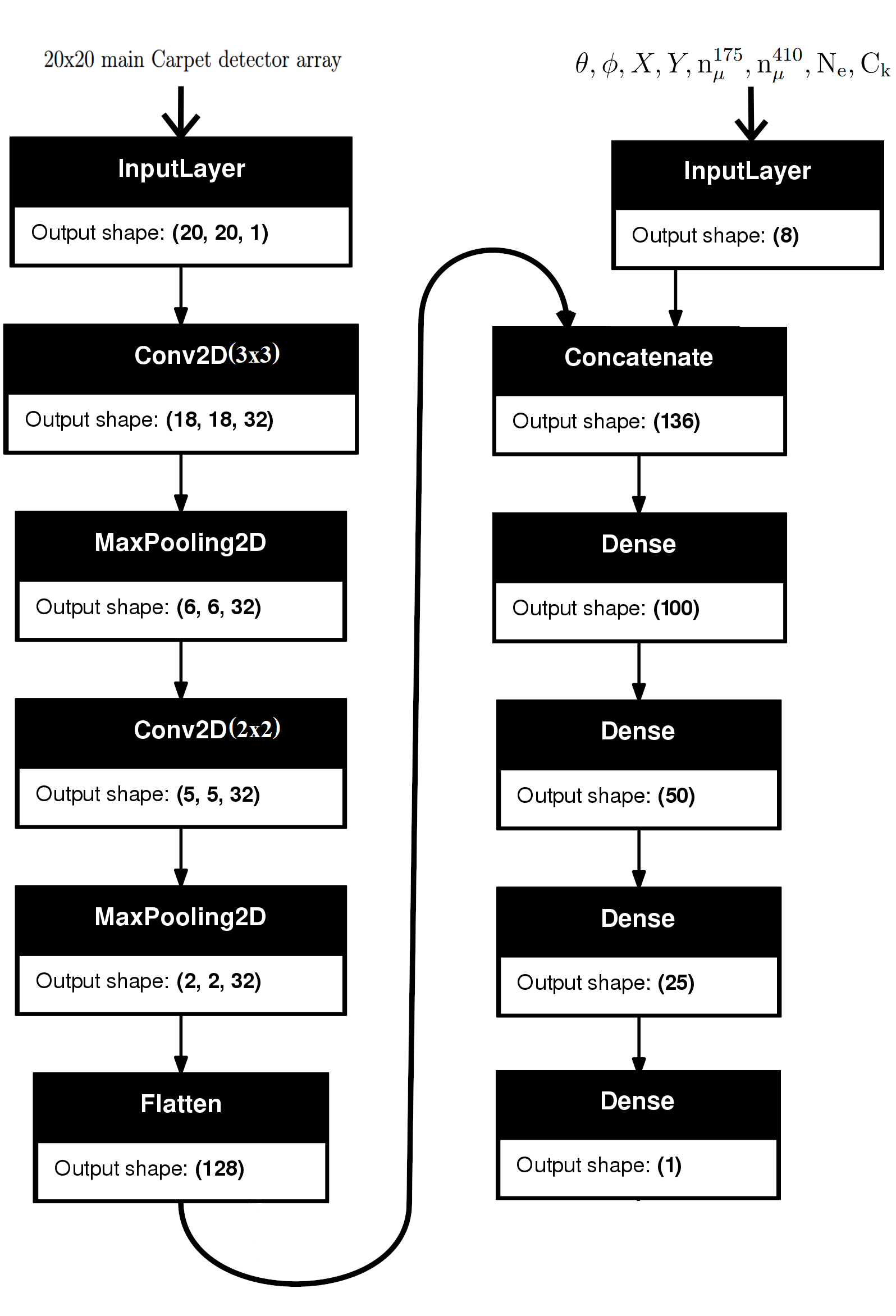}
\caption{
\label{fig:model}
The architecture of the neural network classifier. \textit{Keras} \cite{chollet2015keras} and \textit{Tensorflow} \cite{tensorflow2015-whitepaper} are used as an application programming interface (API) for the neural network.}
\end{figure}

Fig.~\ref{fig:model} shows the architecture of the neural network classifier used in the present paper. The central Carpet array can be seen as a 20-by-20 pixel image, which is submitted as the input to a convolutional part of the network. The sizes of the filters are chosen to be sensitive to typical sub-clusters in the spatial data. Then, the output of this part is merged with the reconstructed data and fed forward to several fully connected layers. The number and size of those layers are chosen based on the performance on the training data. The reconstructed parameters used are arrival angles $\theta$ and $\phi$, shower axis coordinates at ground level $X, Y$, the number of muons $\rm{n}_\mu^{175}$ in the old $175\rm{m}^2$ muon detector part, the number of muons $\rm{n}_\mu^{410}$ in the full $410\rm{m}^2$ muon detector, shower size $N_e$ and the variable $C_k$, introduced in \cite{Conceicao:2022lkc}.

\bibliography{grb}
\end{document}